\documentclass[prb,twocolumn,showpacs]{revtex4}
\usepackage[all]{xy}
\usepackage{amssymb, amsmath}
\usepackage{accents}

\usepackage{graphicx}
\usepackage[caption=false]{subfig}

\begin{document}

\title{Integrability of an extended $d+id$-wave pairing Hamiltonian}
\author{Ian Marquette and Jon Links\\ }
\affiliation{School of Mathematics and Physics, The University of Queensland, Brisbane, QLD 4072, Australia}
\email{i.marquette@uq.edu.au, jrl@maths.uq.edu.au}
\begin{abstract}
We introduce an integrable Hamiltonian which is an extended $d+id$-wave pairing model. The integrability is deduced from a duality relation with the Richardson-Gaudin ($s$-wave) pairing model, and associated to this there exists an exact Bethe ansatz solution. We study this system using the continuum limit approach and solve the corresponding singular integral equation obtained from the Bethe ansatz solution. We also conduct a mean-field analysis and show that results from these two approaches coincide for the ground state in the continuum limit. We identify instances of the integrable system where the excitation spectrum is gapless, and discuss connections to non-integrable models with $d+id$-wave pairing interactions through the mean-field analysis.   
\end{abstract}


\maketitle

\section{Introduction}

The theory of Bardeen, Cooper and Schrieffer\cite{Bar} (BCS) to describe the superconducting properties of metallic materials  is now more than fifty years old. 
An exact Bethe ansatz solution in the case of a reduced BCS Hamiltonian with uniform coupling parameters was obtained by Richardson and Sherman\cite{Ric1,Ric2}. This same system was also studied by Gaudin using an electrostatic analogy and the continuum limit approach.\cite{Gau} Today it is often referred to in the literature as the Richardson model, Richardson-Gaudin model, or $s$-wave model.  Some time later it was shown that this system is integrable,\cite{Cam} and following on from this the integrability and exact solution were unified through approaches involving the quantum inverse scattering method based on the Yang-Baxter equation,\cite{Lin} and Gaudin algebra methods.\cite{dps04} From those developments up to the present day, the $s$-wave pairing model has continued to be the subject of many investigations including the implementation of numerical techniques for solving the associated Bethe ansatz equations,\cite{Dom,Far,Pan} studies using analytic methods,\cite{Cro,gb11,p12} the computation of correlation functions,\cite{dlz05,fcc08,ao12} and quantum dynamics,\cite{fcc09,fcc10} and relations to conformal field theory.\cite{Sed}   

There have also been several searches for examples of integrable, exactly solvable BCS Hamiltonians which go beyond the $s$-wave case. Some examples were found whereby integrability was maintained by the inclusion of additional interaction terms besides the familiar pair-scattering interaction.\cite{ado01,des01,o03,s09} Recently, it was also found that integrability holds in the case of $p+ip$-wave pairing.\cite{Iba,Dun2,Romb} An interesting feature of the $p+ip$-wave pairing case is that the model has a non-trivial ground-state phase diagram, in contrast to the $s$-wave model. A natural question to ask is whether the next in the sequence, a Hamiltonian with $d+id$-wave pairing,\cite{rg00,stf10} also admits an exact solution and if so what is the predicted phase diagram? Within the typical (Lie algebraic) ansatz for the model's eigenfunctions, a recent study\cite{Bir} indicates that the answer is negative regarding the existence of an exact solution. Here we will show that with the inclusion of additional interactions, of the same  quadratic number operator type as those encountered in earlier studies,\cite{ado01,des01,o03,s09} an exact solution can be obtained. Our approach uses the same conserved operators as those of the $s$-wave model, but the Hamiltonian is constructed via a different combination. We examine the continuum limit of the Bethe ansatz equations, and the gap and chemical potential equations from mean-field theory for the ground state, and find that these agree.  The mean-field equations point to the existence of a critical point where the excitation spectrum is gapless. This critical point manifests itself in the Bethe ansatz equations as the point at which all ground-state roots are equal to zero. We finally discuss connections to non-integrable models with $d+id$-wave pairing interactions through the mean-field analysis.

In Sect. 2 we establish integrability of the extended $d+id$-wave pairing Hamiltonian by using a duality transformation applied to the $s$-wave pairing Hamiltonian conserved operators. In Sect. 3 we use the continuum limit of the Bethe ansatz equations to study the the system by solving the corresponding singular integral equation. We also discuss the arc in the complex plane that contains the roots of the Bethe ansatz equations in the continuum limit for uniformly distributed single particle levels (free fermions in two dimensions). In Sect. 4, we use the mean-field approximation to obtain the chemical potential and gap equations. We show that these results coincide with results of Sect. 3 in the continuum limit, and conclude that the results found in Sect. 3 refer to the ground-state of the system. We also discuss connections to critical points for non-integrable cases through the mean-field analysis. Concluding remarks are given in Sect. 4.

\section{The Hamiltonian and integrability} \label{hamint}

We first introduce the Hamiltonian of the singlet-pairing BCS  model with $d+id$-wave symmetry. We take the canonical (i.e. particle number preserving) Hamiltonian whose mean-field approximation leads to the Bogoliubov-de Gennes equations with order  parameter having $d+id$-wave symmetry (e.g. see Read and Green\cite{rg00}, Sato et al.\cite{stf10}) up to quadratic approximation. Letting $c_{{\bf k}\sigma}$, $c_{{\bf  k}\sigma}^{\dagger}$ denote annihilation and creation operators for two-dimensional fermions of mass $m$ with momentum ${\bf k}=(k_{x},k_{y})$,   the Hamiltonian reads    
\begin{align}
H&=\sum_{{\bf k},\sigma}\frac{|\bf{k}|^{2}}{2m}c^\dagger_{{\bf k}\sigma}c_{\bf{k}\sigma} \label{ham1}\\
&\qquad -\tilde{G}\sum_{{\bf k},{\bf k}'}(k_{x}+ik_{y})^{2}(k'_{x}-ik'_{y})^{2}c_{{\bf k}\uparrow}^{\dagger}
c_{-{\bf k}\downarrow}^{\dagger}c_{-{\bf k}'\downarrow}c_{{\bf k}'\uparrow} \nonumber
\end{align}
where $\sigma=\uparrow,\,\downarrow$ denote the spin labels and $G$ is the coupling constant (positive for an attractive interaction). It is more convenient to introduce the following Cooper pair operators  $\tilde{b_{{\bf k}}}^{\dagger}=c_{{\bf k}\uparrow}^{\dagger}c_{-{\bf k}\downarrow}^{\dagger}$,  $\tilde{b_{\bf k}}=c_{-{\bf k}\downarrow}c_{{\bf k}\uparrow}$, and $\tilde{N_{{\bf k}}}=\tilde{b_{{\bf k}}}^{\dagger}\tilde{b_{{\bf k}}}$. For any unpaired fermionic state the action of the pairing interaction is zero and we can decouple the Hilbert space into a product of paired and unpaired fermions states, for which  the action of the Hamiltonian on the space for the unpaired fermions is automatically diagonal in the natural basis. We can therefore exclude unpaired fermions. We consider $m=1$, $\varepsilon_{{\bf k}}=|\textbf{k}|^{2}$ and $k_{x}+ik_{y}=|{\bf k}|\exp(i\phi_{{\bf k}})$ in which case the Hamiltonian takes the form 

\begin{equation}
H=\sum_{{\bf k }}\varepsilon_{{\bf k}}\tilde{N_{{\bf k}}}-\tilde{G}\sum_{{\bf k}, {\bf k}' } \varepsilon_{{\bf k}}\varepsilon_{{\bf k}'}\exp(2i\phi_{{\bf k}})\exp(-2i\phi_{{\bf k}'})\tilde{b_{\bf k}}^{\dagger}\tilde{b_{{\bf k}'}}.
\end{equation}
We next perform the following unitary transformation
\begin{align*}
\tilde{b_{{\bf k}}}^{\dagger}=\exp(-2i\phi_{{\bf k}})b_{{\bf k}}^{\dagger}, \qquad 
\tilde{b_{{\bf k}}}=\exp(2i\phi_{{\bf k}})b_{{\bf k}}.
\end{align*}
Using the integers rather than ${\bf k}$ to enumerate the momentum states, and adding a constant term,
the Hamiltonian takes the form
\begin{align*}
H&=\sum_{j=1}^{L}\varepsilon_{k}N_{j}-\tilde{G}\sum_{j,k=1}^{L}\varepsilon_{j}\varepsilon_{k}b_{j}^{\dagger}b_{k} 
+\frac{\tilde{G}}{2}\sum_{j=1}^L\varepsilon_j^2\\
&=\sum_{j=1}^{L}\varepsilon_{k}N_{j}-\frac{\tilde{G}}{2}\sum_{j,k=1}^{L}\varepsilon_{j}\varepsilon_{k}
\left(b_{j}^{\dagger}b_{k}+b_{j} b_{k}^{\dagger}+ 2\delta_{jk}N_{j}N_{k} \right)
\end{align*}
where the hardcore boson operators satisfy the following commutation relations 
\begin{align*}
 [b_{j},b_{k}^{\dagger}]=\delta_{jk}(1-2b_{j}^{\dagger}b_{j}),\quad [b_{j},b_{k}]=[b_{j}^{\dagger},b_{k}^{\dagger}]=0  
\end{align*}
as well as $(b_{j}^{\dagger})^{2}=0$, $N_j^2=N_j$.
Setting $G=\tilde{G}/2$ we now introduce an extended $d+id$-wave pairing Hamiltonian 
\begin{align}
H=\sum_{j=1}^{L}\varepsilon_{j}N_{j}-G\sum_{j,k}^{L}\varepsilon_{j}\varepsilon_{k}(b_{j}^{\dagger}b_{k}+b_{j}b_{k}^{\dagger}+2N_{j}N_{k}),  \label{nham}
\end{align}
We will establish integrability of (\ref{nham}) using the integrability properties of the $s$-wave model and a duality transformation involving the coupling constant $G$ and the single particle energy levels $\varepsilon_k$. We assume that the $\varepsilon_k>0$ are ordered such that $\varepsilon_j<\varepsilon_k$ whenever $j<k$, but otherwise there are no constraints imposed.

The $s$-wave pairing Hamiltonian is given in terms of hardcore boson operators by
\begin{align}
H=2\sum_{j=1}^{L}\epsilon_{j} N_{j}- G' \sum_{j,k=1}^{L}b_{j}^{\dagger}b_{k}.  \label{ham2}
\end{align}
The integrability of this Hamiltonian was shown by explicit construction\cite{Cam} of a set of commutating operators (integrals of motion).  
These operators read
\begin{equation}
 \tau_{j}=-\frac{2}{G'}(2N_{j}-I) +\sum_{k \neq j}^{L}\frac{\theta_{jk}}{\epsilon_{j}-\epsilon_{k}},  \label{cons1}
\end{equation}
with 
$$\theta= b^{\dagger} \otimes b + b \otimes b^{\dagger}+\frac{1}{2}(2N-I) \otimes (2N -I).$$
The eigenvalues of these operators (excluding the space of unpaired fermions) are given by\cite{Lin,dps04} 
\begin{equation}
\lambda_{j}=\frac{1}{G'}+\frac{1}{2}\sum_{k\neq j}^{L}\frac{1}{\epsilon_{j}-\epsilon_{k}}-\sum_{l=1}^{M}\frac{1}{\epsilon_{j}-v_{k}},\label{ev}
\end{equation}
where the $v_k$ are solutions of the corresponding Bethe ansatz equations 
\begin{equation}
\frac{2}{G'} +\sum_{k=1}^{L}\frac{1}{v_{l}-\epsilon_{k}}=2\sum_{j\neq l}^{M}\frac{1}{v_{l}-v_{j}}.  \label{ba}
\end{equation}
The Hamiltonian given by Eq.(\ref{ham2}) can be written in terms of these commutating operators (up to a additive constant) as 
\begin{align}
H&=-G'\sum_{j=1}^{L}\epsilon_{j}\tau_{j}+\frac{(G')^{3}}{4}\sum_{j,k=1}^{L}\tau_{j}\tau_{k} \nonumber \\ 
&\qquad\qquad +\frac{(G')^{2}}{2}\sum_{j=1}^{L}\tau_{j}-\frac{G'}{2}\sum_{j=1}^{L}C_{j},  \label{comb}
\end{align}
with $C=b^{\dagger}b+bb^{\dagger}+2(2N-I)^{2}$. The energy spectrum of (\ref{ham2}) is given by 
\begin{align*}
 E=2\sum_{j=1}^{M}v_{j}  
\end{align*}
where $M$ represents the total number of Cooper pairs. The above discussion outlines the essential features concerning the integrability and exact solution of the Richardson-Gaudin model (\ref{ham2}). 

We now consider the following change of variables and insert this into the expressions for the integrals of motion, their eigenvalue formulae, and the Bethe ansatz equations:
\begin{equation}
\epsilon \rightarrow \varepsilon^{-1},\quad v \rightarrow y^{-1}.  \label{cv}
\end{equation}
We have from Eqs. (\ref{cons1}), (\ref{ev}), and (\ref{ba}) respectively 
\begin{align}
\tau_{j}&=-\frac{2}{G'} (2N_{j}-I)+\sum_{k \neq j}^{L}\frac{\varepsilon_{j}\varepsilon_{k} \theta_{jk}}{\varepsilon_{k}-\varepsilon_{j}},  \label{ncons} \\
\lambda_{j}&=\frac{1}{G'} +\frac{1}{2}\sum_{k \neq j}^{L}\frac{\varepsilon_{j}\varepsilon_{k}}{\varepsilon_{k}-\varepsilon_{j}}-\sum_{l=1}^{M}\frac{y_{l}\varepsilon_{j}}{y_{l}-\varepsilon_{j}}, \label{ev2} \\
\frac{2}{G' y_{l}}&+\sum_{k=1}^{L}\frac{\varepsilon_{k}}{\varepsilon_{k}-y_{l}}=2\sum_{j\neq l}^{M}\frac{y_{j}}{y_{j}-y_{l}}.  \label{ba2}
\end{align}
Any combination of the conserved operators will commute with the full set of conserved oeprators and can be taken as an integrable Hamiltonian. Instead of the combination given by Eq. (\ref{comb}) that corresponds to the $s$-wave Hamiltonian (\ref{ham2}), we take the following combination, written in terms of the new parameters $\varepsilon_{i}$ given by Eq. (\ref{cv})
\begin{equation}
\tilde{H}=-\frac{G'}{2}\sum_{j=1}^{L}\varepsilon_{j}\tau_{j},  \label{ncomb}
\end{equation}
%
%
Next we implement the following transformation of the coupling constant 
$$4G^{-1}=-G'^{-1}+2\sum_{j=1}^L\varepsilon_{j}.$$ 
We are now in a position to express the Hamiltonian of the extended $d+id$-wave pairing Hamiltonian (\ref{nham}) in terms of the Hamiltonian $\tilde{H}$ given by (\ref{ncomb}):
\begin{align*}
H=G\left(-\frac{4}{G'} \tilde{H}-\frac{2}{G'} \sum_{j=1}^{L}\varepsilon_{j}+\frac{1}{2}\sum_{j=1}^{L}\sum_{k\neq j}^{L}\varepsilon_{j}\varepsilon_{k}-\sum_{j=1}^{L}\varepsilon_{j}^{2}\right).  
\end{align*}
The energy of (\ref{nham}) is found to be given by the roots of the Bethe ansatz equations 
\begin{equation}
\frac{1}{2Gy_{l}}+\sum_{i=1}^{L}\frac{\varepsilon_{i}^2}{y_{l}(y_{l}-\varepsilon_{i})}=2\sum_{j\neq l}^{M}\frac{y_{j}}{y_{l}-y_{j}},\quad l=1,...,M  \label{nba}
\end{equation}
through the expression 
\begin{equation}
E=\sum_{l=1}^{M}y_{l}-2G\sum_{l=1}^{M}\sum_{j \neq l}^{M}y_{j}y_{l}-G\sum_{j=1}^{L}\varepsilon_{j}^{2}.  \label{ne}
\end{equation}
Eqs. (\ref{nba}) and (\ref{ne}) provide the exact solution of the extended $d+id$-wave pairing model (\ref{nham}) on the space with no unpaired fermions. Extending this solution to the full Hilbert space of states is straightforward by omitting the appropriate blocked levels from the summations over $\varepsilon_j$, and by accounting for the contribution of the blocked states to the total energy.

The eigenstates of Hamiltonian are also obtained in the above method. In terms of the solutions of (\ref{nba}) they read  
\begin{align*}
\left|\Psi\right\rangle = \prod_{j=1}^M C(y_j)\left|0\right\rangle,\quad C(y)=\sum_{k=1}^L\frac{\varepsilon_{k}}{y-\varepsilon_{k}}b_{k}^{\dagger}
\end{align*}
where $\left|0\right\rangle$ is the vacuum state.
It can also be shown via a direct algebraic calculation that the following identity holds:
\begin{align*}
[H,C^{M}(0)]=MC^{M-1}(0)Q^{\dagger} (2 G\sum_{l=1}^L\varepsilon_{l}-1).  
\end{align*}
This property is reminiscent of the $p+ip$-wave pairing model.\cite{Iba,Dun2,Romb}  Given any eigenstate $\left|\Psi\right\rangle$ with energy $E$, the {\it dressed} state  $\left|\Psi'\right\rangle=C^M(0)\left|\Psi\right\rangle$ is also an eigenstate with energy $E$ whenever $2 G\sum_{l=1}^L\varepsilon_{l}=1$. In the following analysis it will be seen that states which are dressings of the vacuum state will play a distinguished role.      

\section{Continuum limit and singular integral equation} \label{cl}
In this Section we study the continuum limit of the Bethe ansatz equations, and look to solve the associated singular integral equation.
The Bethe ansatz equations as given by (\ref{nba}) and can be written as
\begin{align*}
 \frac{C}{y_{l}}+\frac{D}{y_{l}^{2}} +\sum_{i=1}^{L}\frac{1}{y_{l}-\varepsilon_{i}}= 2\sum_{j \neq l}^{M}\frac{1}{y_{l}-y_{j}},
 \quad l=1,...,M   
\end{align*}
for $C=2(M-1)-L$ and 
$$D=\frac{1}{2G}-\sum_{i=1}^{L}\varepsilon_{i},$$
The Bethe ansatz equations for $D=0$ would admit an electrostatic analogy as has been used in other studies.\cite{Gau,dps04,gb11,p12,ao12,Dun2,Romb} In that setting the solutions $y_j$ are equilibrium positions of $M$ mobile $+1$ charges subject to electrostatic forces from 
$-{1}/{2}$ charges associated to the  $L$  fixed positions $\varepsilon_{i}$, and a charge $-{C}/{2}$ at the origin. Even though the electrostatic analogy  no longer holds for $D \neq 0$, the equation can still be solved by use of complex analysis techniques. In the continuum limit $G^{-1},M,\,L\rightarrow \infty$ with $x=M/L$ and $g=GL$ held constant, we assume that the roots $y_j$ are distributed according to a density function $r(y)$ with support on an arc $\Gamma$ in the complex plane. Introducing a second density function $\rho(\varepsilon)$ with support on an interval $\Omega=(0,\omega)$ of the real line such that 
$$\int_{\Omega}d\varepsilon \,\rho(\varepsilon)=1$$
we can write
\begin{equation}
\frac{c}{y}+\frac{d}{y^2} - \int_{\Omega}d\varepsilon  \,\frac{\rho(\varepsilon)}{\varepsilon-y}+P\int_{\Gamma}|dy'|\, \frac{2r(y')}{y'-y}=0  , \label{iequa}
\end{equation}
where $c= 2x -1$  and
\begin{equation}
d = \frac{1}{2g}-\int_{\Omega}d\varepsilon \,\varepsilon \rho(\varepsilon) . \label{param}
\end{equation}

We first consider the case $\Gamma$ = $\Gamma_{A}\cup \Gamma_{B} $ where $\Gamma_{A}=(0,\varepsilon_{A})$, $\varepsilon_{A} < \omega $ and $\Gamma_{B}$ is an arc, cutting the real axis at $\varepsilon_{A}$, which is invariant under complex conjugation and has end points $a,b$ such that $a=b^*=\epsilon + i \delta $. In this case Eq. (\ref{iequa}) becomes, $\forall y \in \Gamma_{B}$, 
\begin{align}
\frac{c}{y}+ \frac{d}{y^{2}}+\int_{0}^{\varepsilon_{A}}d\varepsilon \frac{\rho(\varepsilon)}{\varepsilon-y}&-\int_{\varepsilon_{A}}^{\omega}d\varepsilon \frac{ \rho(\varepsilon)}{\varepsilon-y} 
\nonumber \\
&=P\int_{\Gamma_{B}}|dy'| \frac{2r(y')}{y-y'}. \label{sie}
\end{align}
We introduce a function $h(y)$ which is analytic outside of $0,\,\Omega$ and $\Gamma_{B}$, with a branch cut on 
$\Gamma_{B}$ such that $2\pi i r(y)|dy|=(h_{+}(y)-h_{-}(y))dy$ $\forall y \in \Gamma_{B}$. Specifically
\begin{align}
h(y)&= R(y)\left( \int_{0}^{\omega} d\varepsilon \frac{\phi(\varepsilon)}{\varepsilon-y} +\frac{T}{y}+\frac{U}{y^{2}}\right), \label{hrfunc} \\
R(y)&=\sqrt{(y-a)(y-b)}
\end{align}
where $R(\varepsilon)$ is positive on $(\varepsilon_{A},\omega)$ and negative on $(0,\varepsilon_{A})$. Letting $C_B$ denote a closed contour around $\Gamma_B$, the Cauchy principal value of the expression in Eq. (\ref{sie}) can be written as 
\begin{equation}
P\int_{\Gamma_{B}}2|dy'|\frac{r(y')}{y'-y}=\oint_{C_{{B}}}\frac{dy'}{2\pi i}\frac{h(y')}{y'-y}. \label{cpv}
\end{equation}
Using residues to evaluate (\ref{cpv}) for the ansatz (\ref{hrfunc}), and substituting into (\ref{sie}) leads to 
\begin{align*}
&\frac{c}{y}+ \frac{d}{y^{2}}+\int_{0}^{\varepsilon_{A}}d\varepsilon \frac{\rho(\varepsilon)}{\varepsilon-y}-\int_{\varepsilon_{A}}^{\omega}d\varepsilon \frac{ \rho(\varepsilon)}{\varepsilon-y}
\\&\quad
+ \int_{\varepsilon_{A}}^{\omega} d\varepsilon \frac{\phi(\varepsilon)R(\varepsilon)}{\varepsilon-y}-\int_{0}^{\varepsilon_{A}} d\varepsilon \frac{\phi(\varepsilon)|R(\varepsilon)|}{\varepsilon-y}  -\int_{0}^{\omega}d\varepsilon \phi(\varepsilon) \\
&\qquad\quad + {T}  \left(1- \frac{\sqrt{a b}}{y}\right) 
+{U}   \left( \frac{a+b}{2\sqrt{ab}y}-\frac{\sqrt{ab}}{y^{2}}\right)   =0.
\end{align*}
The above equation is satisfied by choosing
\begin{align*}
\phi(\varepsilon)=\frac{\rho(\varepsilon)}{|R(\varepsilon)|} ,\quad U=\frac{d}{\sqrt{ab}} ,\quad T=\frac{c}{\sqrt{ab}}+\frac{d(a+b)}{2ab\sqrt{ab}}, 
\end{align*}
and
\begin{equation}
\int_{0}^{\omega}d\varepsilon\, \phi(\varepsilon)=\frac{c}{\sqrt{ab}}+\frac{d(a+b)}{2ab\sqrt{ab}}  . \label{rel4}
\end{equation}

Given this solution we may now calculate quantities of interest. The first example is the filling fraction $x$:
\begin{align*}
x&= \int_{\Gamma_{A}}|dy|\,r(y)+\int_{\Gamma_{B}}|dy\,|r(y) \\
&= \int_{0}^{\varepsilon_{A}}d\varepsilon \,\rho(\varepsilon)  +\frac{1}{4\pi i} \oint_{C_{{B}}}{dy}\, h(y) \\
&=\frac{1}{2}+\frac{c}{2}+\frac{d}{2\sqrt{ab}}-\frac{1}{2}\int_{0}^{\omega}d\varepsilon \,\varepsilon \phi(\varepsilon). 
\end{align*}
However, as the parameter $c$ also depends on the filling fraction, the $x$ terms cancel and we obtain the following equation 
\begin{equation}
\frac{1}{g}=2\int_{0}^{\omega}d\varepsilon \,\varepsilon \rho(\varepsilon)   + 2\sqrt{a b} \int_{0}^{\omega} d\varepsilon\,\varepsilon \phi(\varepsilon)  . \label{equation1}
\end{equation}
Now combining (\ref{param},\ref{rel4},\ref{equation1}) gives the following equation for the filling fraction
\begin{equation}
x=\frac{1}{2}+\frac{\sqrt{ab}}{2}\int_{0}^{\omega}d\varepsilon \,\phi(\varepsilon) -\frac{(a+b)}{4\sqrt{a b}}\int_{0}^{\omega}d\varepsilon \,\varepsilon \phi(\varepsilon) . \label{equation2}
\end{equation}
Defining the intensive energy $e$ as
$$e=\lim_{L\rightarrow \infty} \frac{E}{L}$$ 
we have from (\ref{ne}) that $e={\mathcal I}-2g{\mathcal I}^2$ where
\begin{align}
{\mathcal I}&=\int_{\Gamma_{A}}|dy|\,y r(y)+\int_{\Gamma_{B}}|dy|\,y r(y) \nonumber \\
&=\int_{0}^{\varepsilon_{A}}d\varepsilon \,\varepsilon \rho(\varepsilon)  + \oint_{C_{{B}}}\frac{dy}{4\pi i}y h(y) 
\nonumber \\
&=-\frac{d(\sqrt{a}-\sqrt{b})^{2}}{4\sqrt{ab}}  +  \frac{1}{2}\int_{0}^{\omega} d\varepsilon \,\varepsilon \rho(\varepsilon) 
\nonumber\\
&\qquad -\frac{1}{4}\int_0^\omega d\varepsilon\,\varepsilon\phi(\varepsilon)(2\varepsilon-a-b) \nonumber \\
&=\frac{1}{2}\int_0^\omega d\varepsilon\,\varepsilon\rho(\varepsilon)-\frac{1}{2}\int_0^\omega d\varepsilon\,\varepsilon^2\phi(\varepsilon) \nonumber \\
&\qquad +\frac{\sqrt{ab}}{2}\int_0^\omega d\varepsilon\,\varepsilon\phi(\varepsilon)
\label{equation3}
\end{align}
where in the last step we use (\ref{param},\ref{equation1}).

Repeating the above calculations for $\Gamma=\Gamma_B$ cutting the real axis at $\varepsilon_{A}<0$, again invariant under complex conjugation and having end points $a,b$ such that $a=b^*=\epsilon + i \delta $  leads to exactly the same Eqs. (\ref{equation1},\ref{equation2},\ref{equation3}). 

\subsection{Equation of the arc}

We now specialise to the case of energy levels $\varepsilon$ uniformly distributed in the interval $[0,1]$ (i.e. taking $\omega=1$), such that $\rho(\varepsilon)=1$. This case corresponds to the distribution of
two-dimensional free fermions. Integrating Eqs. (\ref{equation1},\ref{equation2}) leads to 
\begin{align*}
\frac{1}{g}&=1+2\sqrt{\epsilon^{2}+\delta^{2}}(\sqrt{(\epsilon-1)^{2}+\delta^{2}}-\sqrt{\epsilon^{2}+\delta^{2}}) \\
&\qquad\quad+2\epsilon\sqrt{\epsilon^{2}+\delta^{2}} \ln \left|\frac{ \sqrt{(1-\epsilon)^{2}+\delta^{2}}+1-\epsilon}{\sqrt{\epsilon^{2}+\delta^{2}}-\epsilon}\right| \\
x&=\frac{1}{2}+\frac{\epsilon}{2}\left(1-\sqrt{\frac{{(\epsilon-1)^{2}+\delta^{2}}}{{\epsilon^{2}+\delta^{2}}}} \right)  \\
&\qquad\quad+\frac{\delta^2}{2\sqrt{\epsilon^2+\delta^2}} \ln \left|\frac{ \sqrt{(\epsilon-1)^{2}+\delta^{2}}+1-\epsilon}{\sqrt{\epsilon^{2}+\delta^{2}}-\epsilon}\right| .
\end{align*}
For a given value of the coupling constant $g$ and filling fraction $x$ we can numerically obtain the end points $\epsilon\pm i\delta$. From these end points we can numerically evaluate the following integral that gives the equation of $\Gamma_B$ in the complex plane:
\begin{equation}
{\rm Re}\left(\int_{a}^{y}dy'h(y')\right)=0,\quad y \in \Gamma_{B}. \label{arcopen}
\end{equation}

\begin{figure} 
\centering
\subfloat[]{\includegraphics[width=6.8cm]{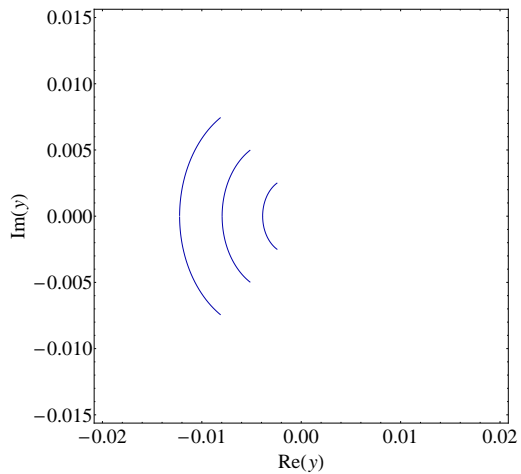}}\\
\subfloat[]{\includegraphics[width=6.8cm]{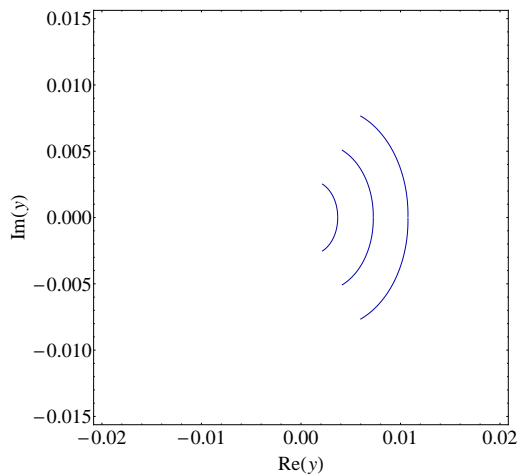}}
\caption{Arcs representing solutions of the Bethe ansatz equations in the continuum limit. Results are shown in the case of $\rho(\varepsilon)=1$, $x=1/2$ for (a) the arc $\Gamma$ with $g={203}/{200}$, $g={202}/{200}$ and $g={201}/{200}$; (b) the arc $\Gamma_B$ with $g={199}/{200}$,  $g={198}/{200}$ and  $g={197}/{200}$. The limiting behaviour indicates that the arc $\Gamma$ contracts to a point at the origin when $g=1$. This indicates that the corresponding eigenstate at $g=1$ is the state obtained by dressing the vacuum.}
\label{fig1}
\end{figure}

For the uniform distribution $\rho(\varepsilon)=1$ it is easily verified that Eq. (\ref{equation1}) is satisfied by the choice $g=1,\epsilon=\delta=0$ giving $e=\mathcal{I}=0$. We recognise the corresponding eigenstate as the dressing of the vaccum as discussed in Sect. \ref{hamint}. Fig. 1 shows the limiting behaviour of the arc $\Gamma$ as $g\rightarrow 1$, confirming this conclusion. We highlight that this scenario is very different from the case of the $p+ip$-wave model. In that model  roots collapse at the origin when the coupling crosses the so-called {\it Moore-Read} line for any finite number of particles, but the limiting behaviour of the arc in the continuum limit is a closed curve, of finite length, encircling the origin.\cite{Dun2,Romb}

In the above calculations we have obtained a unique solution of the Bethe ansatz equations in the continuum limit under the assumption that the roots become dense on a connected curve in the complex plane. Now we will argue that this solution corresponds to the ground state of the Hamiltonian, by making comparison between the exact solution and results obtained from a mean-field approximation of the Hamiltonian. Using a mean-field approximation will also allow us to make some comparison between the properties of the $d+id$-wave pairing Hamiltonian (\ref{ham1}) and those of the exactly solvable extended $d+id$-wave pairing Hamiltonian (\ref{nham}). 

\section{Mean-field approximation}

 We consider a Hamiltonian with the compactly written form 
\begin{equation}
\mathcal{H}=H_{0}-GQ^{\dagger}Q-GQQ^{\dagger}-2F H_{0}^{2}, \label{hamilonian}
\end{equation}
where $\displaystyle H_{0}=\sum_{k=1}^{L}\varepsilon_{k}N_{k}$, $\displaystyle Q=\sum_{k=1}^{L}\varepsilon_{k}b_{k}$ and $F,\,G\geq0$ are the two coupling parameters. 
Using a mean-field approximation the Hamiltonian given by Eq. (\ref{hamilonian}) becomes
\begin{align}
\mathcal{H}&=H_{0}-\frac{1}{2}\check{\Delta}^*Q-\frac{1}{2}\check{\Delta}Q^{\dagger}+\frac{\Delta^{2}}{8G} \nonumber \\
&\qquad\quad -4F\Gamma H_{0}+2F \Gamma^{2}-\mu(N- M), \label{hamiltonianmean}
\end{align}
where $\check{\Delta}=4G\left\langle Q \right\rangle$, $\Gamma=\left\langle H_{0} \right\rangle$, $\Delta=|\check{\Delta}|$, $N$ is the Cooper pair number operator, $M=\langle N \rangle$, and $\mu$ is a Lagrange multiplier which is introduced since the mean-field approximation does not conserve particle number. We will assume that the total fermion number is even, so all fermions are paired in the ground state. Setting $F=0$ gives a mean-field approximation for (\ref{ham1}) (up to a canonical transformation), while setting $F=G$ gives a mean-field approximation for (\ref{nham}). Keeping $F$ as a free variable allows us to interpolate between these two cases.

Setting $s=1-4F\Gamma$ and 
\begin{align}
\mathcal{E}(\varepsilon_{j})=\sqrt{(s \varepsilon_{j}-\mu)^{2}+\varepsilon_{j}^{2}\Delta^{2}} 
\label{spectra}
\end{align}
the ground-state energy is found to be  
\begin{equation}
E=\frac{1}{2}\sum_{j=1}^L(s \varepsilon_{j}-\mu)-\frac{1}{2}\sum_{j=1}^L\mathcal{E}(\varepsilon_{j})+\frac{\Delta^{2}}{8G}+2F\Gamma^{2}+\mu M \label{energymean}
\end{equation}
associated to the mean-field ground state  
$$\left|\Psi_{\rm mf}\right\rangle=\prod_{j=1}^L(u_{j}I+v_{j}b_{j}^{\dagger})  \left|0\right\rangle$$
where
\begin{align*}
|u_{j}|^{2}=\frac{1}{2}\left(1+\frac{s\varepsilon_{j}-\mu}{\mathcal{E}(\varepsilon_{j})}\right),\quad |v_{j}|^{2}=\frac{1}{2}\left(1-\frac{s\varepsilon_{j}-\mu}{\mathcal{E}(\varepsilon_{j})}\right). 
\end{align*}
The elementary excitation energies are given by $\mathcal{E}(\varepsilon_{j})/2$. It is seen that the excitation spectrum becomes gapless in the limit $\varepsilon_1\rightarrow 0$ when $\mu=0$.    
 
Through use of the Hellmann-Feynman theorem we may take partial derivatives of (\ref{hamiltonianmean}) and (\ref{energymean}) to generate the following constraint equations:
\begin{align}
\frac{1}{2G}&=\sum_{j=1}^L\frac{\varepsilon_{j}^{2}}{\mathcal{E}(\varepsilon_{j})} , \label{hf1} \\
L-2M&=\sum_{j=1}^L\frac{s\varepsilon_{j}-\mu}{\mathcal{E}(\varepsilon_{j})}, \label{hf2} \\
\sum_{j=1}^L\varepsilon_{j}-2\Gamma&=\sum_{j=1}^L\frac{s\varepsilon_{j}^{2}-{\mu} \varepsilon_{j}}{\mathcal{E}(\varepsilon_{j})}. \label{hf3}
\end{align}
Combining Eqs. (\ref{hf1},\ref{hf3}) gives 
\begin{equation}
\frac{1+4(G-F)}{2G}\Gamma=\sum_{j=1}^L\varepsilon_{j}+\mu \sum_{j=1}^L\frac{\varepsilon_{j}}{{\mathcal{E}(\varepsilon_{j})}}.
\label{gamma}
\end{equation}
Equivalently, the explicit expression for $s$ in term of $F$ and $\Gamma$  leads to 
\begin{equation}
\frac{1-s}{2F}+\frac{s}{2G}=\sum_{j=1}^L\varepsilon_{j}+\mu \sum_{j=1}^L\frac{\varepsilon_{j}}{{\mathcal{E}(\varepsilon_{j})}}. \label{hf1mod0}
\end{equation}
Using (\ref{hf1},\ref{hf2},\ref{hf3}) 
the energy expression (\ref{energymean}) may be rewritten as 
\begin{align}
E=\Gamma-2F\Gamma^2-\frac{\Delta^2}{8G} \label{e1}
\end{align}
which is consistent with (\ref{hamilonian}).

First we consider the integrable case $F=G$ for which (\ref{hf1mod0}) reduces to 
\begin{equation}
\frac{1}{2G}=\sum_{j=1}^L\varepsilon_{j}+\mu \sum_{j=1}^L\frac{\varepsilon_{j}}{{\mathcal{E}(\varepsilon_{j})}}. \label{hf1mod}
\end{equation}
Expressing (\ref{spectra}) as 
$\mathcal{E}(\varepsilon)=\sqrt{s^2+\Delta^2}\sqrt{(\varepsilon-a)(\varepsilon-b)}$
we have 
$$a=b^*= \mu\left(\frac{s + i \Delta }{s^{2}+\Delta^{2}}\right).$$
Identifying 
$$\epsilon=\frac{s\mu}{s^2+\Delta^2},\qquad \delta=\frac{\Delta\mu}{s^2+\Delta^2},$$
in the continuum limit we find that (\ref{hf1mod},\ref{hf2}) and $e$ calculated through (\ref{e1}) coincide with the exact solution expressions (\ref{equation1},\ref{equation2}) and $e$ calculated through (\ref{equation3}) respectively. This gives confidence that the mean-field approximation is valid when $F=G$, while simultaneously showing that the solution found in the continuum limit in Sect. \ref{cl} does correspond to the ground state of the system.
Gapless excitations in the mean-field analysis arise when $\mu=0$ which, from (\ref{hf1mod}), corresponds to   
\begin{align*}
\frac{1}{2G}=\sum_{j=1}^L\varepsilon_{j}
\end{align*}
independent of $x$. The above relation is exactly the condition for dressing states, presented in Sect. \ref{hamint}. 

Finally we consider the mean-field equations for general $F\neq G$ when $\mu=0$. Here we deduce from (\ref{hf1},\ref{hf2},\ref{hf3}) that the excitation spectrum is gapless when 
\begin{align*}
x=\frac{\displaystyle 2G\sum_{j=1}^L \varepsilon_j-1}{\displaystyle 4(G-F)\sum_{j=1}^L\varepsilon_j}
\end{align*}
Qualitatively, the $d+id$-wave case where $F=0$ and the extended $d+id$-wave case with $F=G$ are very similar at the level of the mean-field analysis. For fixed $x$, in both instances there is at most one value of the coupling $G$ at which the spectrum may be gapless. One significant difference however is that gapless excitations for the case $F=0$ can only occur when $x<1/2$, whereas for $F=G$ gapless excitations occur for all filling fractions.  

For non-zero $\varepsilon_j$ we find when $\mu=0$
\begin{align*}
|u_{j}|^{2}=\frac{1}{2}\left(1+\frac{s}{\sqrt{s^2+\Delta^2}}\right),\,\, |v_{j}|^{2}=\frac{1}{2}\left(1-\frac{s}{\sqrt{s^2+\Delta^2}}\right) 
\end{align*}
independent of $j$. Alternatively setting $\varepsilon_1 =0$ and keeping $\mu\neq 0$ we obtain
\begin{align*}
|u_{1}|^{2}=\frac{1}{2}\left(1+\frac{\mu}{|\mu|}\right),\,\, |v_{1}|^{2}=\frac{1}{2}\left(1-\frac{\mu}{|\mu|}\right) 
\end{align*}
which exhibits a discontinuity as $\mu\rightarrow 0$. The case $F=0$ was studied by Read and Green\cite{rg00} who suggested that the $\mu=0$ limit is identified with a critical point for which the ground state is the {\it Haldane-Rezayi} state.\cite{hr88}
However the situation is much the same for generic $F$. Let $P_M$ denote the projection operator onto the subpsace for $M$ Cooper pairs. If we take the projected mean-field ground state at $\mu=0$ (with $\varepsilon_1\neq 0$) we obtain
\begin{align}
\left|\Psi_M\right\rangle &=  P_M \left|\Psi_{\rm mf}\right\rangle 
= C \left(\sum_{j=1}^L b_j^\dagger \right)^M\left|0\right\rangle \label{dressed}
\end{align}
for some normalisation constant $C$. Eq. (\ref{dressed}) is  exactly the dressing of the vacuum state that was discussed in Sect. \ref{hamint}, and appeared in Sect. \ref{cl} in the context of the Bethe ansatz solution. Thus the projected mean-field ground state (\ref{dressed}) for the critical point at which the excitation spectrum is gapless ($\mu=0$) is seen to have a universal character in that it is independent of both $F$ and $G$.

\section{Conclusion}

Using the conserved operators and Bethe ansatz equations of the Richardson-Gaudin pairing Hamiltonian we obtained an  integrable, exactly solvable, extended $d+id$-wave system. This model is curious in that the energy expression depends quadratically on the roots of the Bethe ansatz equations, and also in that the Bethe ansatz equations involve terms which do not allow for an electrostatic analogy. Nonetheless we were still able to study this system by solving the singular integral equation obtained from the Bethe ansatz equations in the continuum limit. We also showed that these results and those obtained in a mean-field approximation coincide. For a general Hamiltonian with two coupling parameters (which includes the integrable case) it was found that the projected mean-field ground-state is universal at the critical point of gapless excitations. This suggests that results from the exactly solvable extended $d+id$-wave system may prove fruitful in understanding the $d+id$-wave system. A precedent for such an approach has been set in the different context of spin ladder systems, whereby it has been shown that extended exactly solvable Hamiltonians can provide extremely useful tools for making comparisons between theoretical calculations and experimental data.\cite{bgot07}

\section*{Acknowledgements}
This work was supported by the Australian Research Council through Discovery Project DP110101414.



\end{document}